# Degressive Representation of Member States in the European Parliament 2019–2024

## Friedrich Pukelsheim and Geoffrey Grimmett

Revision 2018-02-21

*Primary law of the European Union demands that the allocation of the seats of the European Parliament between the Member States must obey the principle of degressive proportionality. The principle embodies the political aim that the more populous states agree to be underrepresented in order to allow the less populous states to be better represented. This paper reviews four allocation methods achieving this goal: the Cambridge Compromise, the Power Compromise, the Modified Cambridge Compromise, and the 0.5-DPL Method. After a year of committee deliberations, Parliament decreed on 7 February 2018 an allocation of seats for the 2019 elections that realizes degressive proportionality, but otherwise lacks methodological grounding. The allocation emerged from haggling and bargaining behind closed doors.*

## 1. Introduction

In the past the seats of the European Parliament (EP) were allocated between the Member States following a strategy of an ever enlarging parliament. Whenever a state acceded to the European Union, its seats were created from scratch and added to the extant total. The generous practice of creating new seats evaded the delicate task of transferring seats from one Member State to another one. However, the strategy has come to an end due to the Treaty of Lisbon. The Treaty limits the size of the EP to at most 751 seats.

With a parliament of limited size future allocations of EP seats will make seat transfers between Member States unavoidable. Moreover the current issue is not the accession of a new state, but the secession of an old state: the Brexit. This secession raises the question of what to do with the 73 UK seats that will become vacant upon Brexit.

The situation forces the EP to review its approach to the allocation of seats between the Member States of the Union. To this end the Constitutional Affairs Committee of the EP (AFCO, from French: *affaires constitutionnels*) conducted a workshop entitled 'The Composition of the European Parliament' in January 2017; the briefings are published in Directorate-General (2017).

The starting point for the workshop was the European Council (2013) decision for establishing the composition of the EP. Art. 4 says that the decision shall be revised before the end of 2016, with the aim of establishing a system to allocate the seats between Member States in an *objective, fair, durable and transparent way*:

> This Decision shall be revised sufficiently far in advance of the beginning of the 2019–2024 parliamentary term on the basis of an initiative of the European Parliament presented before the end of 2016 with the aim of establishing a system which in future will make it possible, before each fresh election to the European Parliament, to allocate the seats between Member States in an objective, fair, durable and transparent way, translating the principle of degressive proportionality as laid down in Article 1, taking account of any change in their number and demographic trends in their population, as duly ascertained thus respecting the overall balance of the institutional system as laid down in the Treaties.

Part 1 of this paper details the legal ramifications promoting, and delimiting, degressivity. Part 2 reviews the four allocation methods presented to AFCO during the January 2017 workshop. Part 3 comments on the final act of decision-making. Another two methods entered the stage, though neither met with AFCO's approval.

In the end AFCO decreed a composition of the 2019 EP elections that emerged from deliberations behind closed doors. It satisfies degressivity, but loses the link to direct representation of the Union citizens. In their debates neither AFCO nor the EP cared to enlighten the public how they arrived at their conclusion. Their motives remain obscure. Parliament missed this opportunity to proceed from the dark ages to an era of enlightenment.

## 2. Degressive Representation

The oxymoron of *degressive proportionality* has an enthusiastic tradition in the debates of the EP. However, the term calls for an explanation. Just as one may have degressive taxation, proportional taxation, or progressive taxation, one may also have degressive representation, proportional representation, or progressive representation. Taken literally *degressive proportionality* is a paradoxical compound. Instead we prefer to speak of *degressive representation*, which is the concept really meant, or *degressivity*, for short.

A parliamentary resolution of 2007 (European Parliament 2008) interprets degressivity to be a manifestation of solidarity:

> The more populous states agree to be underrepresented in order to allow the less populous states to be represented better.

The resolution includes an attempt of a formal specification of degressivity which has since been recognized as a potential contradiction. Meanwhile the abstract principle has been given a concrete specification capable of practical implementation. The specification is part of the secondary Union law (requirement 10 below).

Before turning to secondary law, we begin our discussion of degressivity and the other ramifications of EP elections by listing the requirements of primary Union law.

### 2.1. *Requirements of Primary Union Law*

The Union's primary law is set forth in the Treaty on European Union (TEU, Lisbon Treaty); see European Union (2012). It lays conditions upon possible methods of allocating the seats of the EP between the Member States. Of particular relevance are the following requirements which we rearrange and paraphrase to ease cross-referencing in this paper:

1. Citizens are directly represented in the EP (Art. 10(2) TEU).
2. The EP shall be composed of representatives of the Union's citizens (Art. 14(2) TEU).
3. Representation of citizens shall be degressively proportional (Art. 14(2) TEU).
4. The size of the EP shall not exceed 751 seats (Art. 14(2) TEU).
5. Every Member State shall be allocated at least 6 seats (Art. 14(2) TEU).
6. Every Member State shall be allocated at most 96 seats (Art. 14(2) TEU).

There is a tension between the principles of direct representation (requirement 1) and of degressive representation (requirement 3), each of which is stipulated by primary Union law. Requirement 1 supports an allocation proportional to population no matter how populous a state is. Requirement 3 favours an allocation giving some priority to smaller states. The composition that is eventually realized must pay due attention to both principles, and strike a gentle balance between them.

There is a potential ambiguity in the term *Member State* over whether it refers to government or to people. When Member State is interpreted to mean government, the appropriate representative bodies are the European Council and the Council, by Art. 10(2) TEU, rather than the EP. Within the context of the composition of the EP, the term Member State means people; that is, a Member State's citizenry.

Moreover the notion of degressivity is sensitive to the meanings of the term *citizens* in requirements 2 and 3. The meanings differ significantly even though both requirements appear in the same section of Art. 14 TEU. Reference to *Union citizens* (requirement 2) promises to place all citizens on an equal footing. However, the principle of degressivity (requirement 3) discriminates between citizens according to Member States. The citizens of more populous Member States agree to be underrepresented in order to allow the citizens of less populous Member States to be represented better.

## 2.2. *Requirements of Secondary Union Law*

The extended deliberations of the EP on its composition have led to specifications that have found their way into Art. 1 of the European Council (2013) decision:

> 7. Any more populous Member State shall be allocated at least as many seats as any less populous Member State.
> 8. The least populous Member State shall be allocated 6 seats.
> 9. The most populous Member State shall be allocated 96 seats.
> 10. The principle of degressive proportionality shall require decreasing representation ratios when passing from a more populous Member State to a less populous Member State, where the representation ratio of a Member State is defined to be the ratio of its population figure relative to its number of seats before rounding.

Requirements 8 and 9, as reasoned by Council, allow to reflect as closely as possible the spectrum of the Member States' population figures. This reasoning can be met with the current data, but is invalid in general. If Germany's population were drastically smaller, or if a state of the size of Germany (like Turkey) were acceding, 96 seats might be too many even for the most populous Member State

Requirement 10 provides an operational specification of degressivity requiring that the representation ratios are decreasing when passing from a larger state to a smaller state. Our Tables 1 and 2 below exhibit columns whose labels include "RR" – short for representation ratio – in witness of degressivity as thus specified.

## 2.3. *Population Figures*

Available population figures are those used for the qualified majority voting (QMV) rule in the Council of Ministers. Presumably everybody would endorse the aim that each individual who qualifies (in some way) as a Union citizen shall be counted at least once and at most once. That is, he or she is counted exactly once. This modest aim is not achieved easily, considering that the data are gathered by a host of domestic statistical offices before being communicated to EuroStat. To this end it seems appropriate to continue to base all population figures on the internationally (UN) approved notion of *total resident population*. These are the population figures collected by EuroStat for the application of Council's QMV rule.

Since the Council and the EP are constitutional organs with joint governance responsibility, the two institutions ought to employ the same population data. The population figures for Council's QMV rule during the calendar year 2017 are published by European Council (2016). They provide the input data for Table 1 (column "QMV2017") as well as for Table 2 (which table, for lack of space, shows output only).

## 3. The Four Workshop Proposals

During the January 2017 AFCO workshop four allocation methods were presented which are here reviewed only briefly; for proper details see Directorate General (2017). Regrettably none of these methods met with acceptance by AFCO.

### *3.1. Cambridge Compromise*

The *Cambridge Compromise* of Grimmett et al. (2011) may be paraphrased as follows:

> Every Member State is assigned a common number of base seats. The remaining seats are allocated proportionately to population figures, using the divisor method with upward rounding and subject to a maximum allocation of 96 seats.

The Cambridge Compromise achieves degressivity without distorting the meaning of "citizens" beyond the minimum. It does so in each of its two stages. The first stage of assigning base seats treats all Member States alike. This is extremely degressive since it neglects population figures entirely. The second stage of proportional allocation of the remaining seats embodies a mild form of degressivity through the use of *upward* rounding. Upward rounding is known to introduce a slight bias in favour of the less populous states and at the expense of more populous states (Pukelsheim 2017; Chap. 7). This type of bias reinforces degressive effects.

The Cambridge Compromise results in an increased bunching of Member States near the maximum of 96 seats (as permitted by requirement 9). This does not threaten degressivity, yet it disadvantages capped Member State relative to other large states.

Typically, when compared to the 2014 status quo seat allocation, Member States in the top and bottom thirds would gain seats (or stay as is), while middle-sized Member States would lose seats (or stay as is). The loss of seats constitutes a major political obstacle that opposes the suitability of the Cambridge Compromise for EU-27 in its present configuration.

### *3.2. Power Compromise*

The *Power Compromise* of Grimmett, Oelbermann and Pukelsheim (2012) is a variant of the Cambridge Compromise operating on power-adjusted population units rather than on original population figures. It may be worded as follows:

> Every Member State is assigned a common number of base seats. The remaining seats are allocated proportionately to adjusted population units (that is, the population figures raised to a common power) using the divisor method with upward rounding. The number of base seats, the power, and the divisor are determined so that the least populous state is allocated 6 seats, the most populous state is allocated just 96 seats, and the preordained EP seat total is fully utilized.

The Power Compromise is complicated by involving an additional parameter: the power. Generally there is a range of powers that guarantee 96 seats for the most populous state; it can be shown that in such cases the smallest power achieves the highest level of degressivity. An efficient algorithm to determine the power parameter is described in Grimmett, Oelbermann and Pukelsheim (2012) or in Pukelsheim (2017: Sect. 12.9).

It transpires that, with an EP size of 723 seats or more, no Member State has to relinquish any of its 2014 seats. That is, a total of 723 seats is the minimum EP size for which the Power Compromise realizes a *no-loss seat allocation* such that every Member State meets or exceeds its 2014 seat contingent.

However, the Power Compromise achieves degressivity by interpreting the term "citizens" in a rather broad sense. The method replaces lucent population figures, which count concrete citizens, by obscure population units, which measure abstract units. For example

Malta's population of 434 403 citizens would be transformed to 38 336 population units. Does this mean that less than ten percent of the citizenry is accounted for? Or less than ten per cent of each citizen? Neither interpretation seems profitable; the interim power-adjustments remain dubious. Their justification lies in the final result which thereby achieves a higher degree of degressivity.

The Cambridge Compromise may be viewed as prioritizing direct representation over degressivity. In contrast, the Power Compromise allows greater degressivity, but at some cost to direct representation. The two methods yield seat allocations that become increasingly identical as the power parameter moves closer to unity. They coincide when the power equals unity, and this could occur in the future. This possibility of future coincidence of the two methods mitigates the marginal disregard by the Power Compromise of the principle of direct representation.

### 3.3. *Modified Cambridge Compromise*

The procedure proposed by Słomczyński and Życzkowski (2012), called *Modified Cambridge Compromise*, is akin to the Power Compromise. It relies on *population transformations* that depend on a power parameter *d*. The transformation of a state with population *p* is defined to be

$$\frac{96\,(p^d - 434\,404^d) + 6\,(82\,064\,489^d - p^d)}{82\,064\,489^d - 434\,404^d}.$$

By way of standardization the transformation involves the smallest and largest population figures: 434 404 of Malta and 82 064 489 of Germany, and the smallest and largest admissible seat numbers: 6 and 96. The procedure converts the transformed quantities into whole numbers of seats using the rule of standard rounding. The power parameter is determined so that the sum of all seat numbers exhausts the preordained seat total.

Specifically, the authors find the minimum EP size for a no-loss seat allocation to be 721 seats. This EP size conforms well to the minimum size of 723 seats that are required for a no-loss seat allocation when using the Power Compromise.

### 3.4. *0.5-DPL Method*

Yet another procedure that was presented in the workshop is called the *0.5-DPL Method*. The name includes DP to point to degressive proportionality, while L is reminiscent of the limitations due to the minimum requirement of 6 seats and the maximum capping of 96 seats. The approach originates from the work of Ramírez, Palomares and Márquez (2006).

The 0.5-DPL Method is based on *adjusted quotas*. The adjusted quota is a sum of two terms. The first term is 0.5 times the population figure divided by the sum of all population figures. The second term is 0.5 times the square-root population divided by the sum of all square-root populations. With reference to the adjusted quotas, the divisor method with upward rounding is used for the allocation of the given seat total.

The 0.5-DPL Method fails to realize a no-loss allocation, but not as violently as the Cambridge Compromise. The state that continues to be deficient with regard to the 2014 status quo is Lithuania. Compared to its 11 seats in 2014, Lithuania falls short of at least two seats up to EP size 729, while EP sizes from 730 through 751 feature a one-seat deficit.

## 4. AFCO Deliberations and EP Resolution

The January 2017 workshop on the composition of the EP was followed by a period of silence. The two rapporteurs on the dossier, Danuta Maria Hübner (EPP - PL, AFCO president) and Pedro Silva Pereira (S&D - PT), worked behind the scenes to explore feasible options with the shadow rapporteurs from the Political Groups in the EP. In April 2017 the two rapporteurs issued a working paper, and in September 2017 a draft report. Consideration of the draft report in AFCO sessions was repeatedly postponed, reflecting the challenge of reaching a compromise able to attract majority support. Eventually, in January 2018, the AFCO adopted the report Hübner and Silva Pereira (2018). The proposed composition of the EP for the legislative period 2019−2024 is shown in column "2019" of Table 2. In the sequel we summarise the route taken to this composition.

### *4.1. A Midsummer's Wish List*

During the summer of 2017 AFCO signalled interest in a model devised to meet the following criteria:

- No Member State loses any seats.
- There is respect for degressivity (as specified in requirement 10 above).
- A parliament should have either 700 and 710 members.

The participants of the January 2017 workshop responded with two notes. Grimmett et al. (2017) suggested a no-loss variant of the Cambridge Compromise (see Sect. 4.5 below), and Ramírez González (2017) proposed a variant of the 0.5-DPL procedure, namely the so-called FPS method (Sect. 4.6). Since the eventual AFCO resolution adopted a parliament of 705 seats, we adopt similarly a house size of 705 in the subsequent analysis of this paper.

### *4.2. The Impact of Brexit*

The impending Brexit played an unsurprisingly prominent role in the deliberations of the AFCO. Indeed, there is evidence in the televised sessions of AFCO that speculation about Brexit featured at least as prominently in the discussion as did the issue of fair representation of half a billion Union citizens in the EP.

The AFCO approached the issue of Brexit as follows. In the (unlikely) event that the United Kingdom continues to be represented in the 2019–2024 EP, the current 2014 composition will be maintained. Otherwise, the 2019 composition will be implemented. If the UK exodus takes place during the 2019–2024 legislation period, parliament will promptly switch from the 2014 composition to the 2019 composition. For such a switch to be uncontroversial, it will be necessary that no Member State loses seats. With a no-loss 2019 allocation, the switch may be enacted by the simple mechanism of inviting new deputies into office, thus avoiding the collateral damage of eviction of incumbent Members of Parliament.

The imperative of planning for Brexit provides justification for the *no loss* requirement, namely that every Member State has at least as many seats in the 2019 composition as in the existing 2014 composition. Such a no-loss warranty has been warmly welcomed by the AFCO, since it counters political damage and disputation.

### *4.3. Non-Degressivity of the 2014 Composition*

Unfortunately, the 2014 composition violates degressivity, and not just at one place but at no fewer than ten (see column "RR2014" of Table 1). It would be absurd for the AFCO to postulate degressivity as a *conditio sine qua non*, and then to adopt a "pragmatic" solution with multiple violations.

This awkward failure was recognised by the AFCO, who agreed to remedy it by a judicious scattering of seats from the pot liberated by the departing United Kingdom, thereby raising pragmatism to a new level. On close inspection of the status quo 2014 composition, it is easily seen that the seat contingents of ten Member States need to be raised by a total of 16 seats in order to realize degressivity. The resulting seat allocation is exhibited in column "2014DP" of Table 1, and it utilizes 694 seats.

These 694 seats now take the role of minimum requirements, and each midsummer scenario is now feasible: a parliament of 700 members, as well as a parliament of 710 members. With no public hint of explanation, the AFCO has settled on 705.

### 4.4. *No-Loss Cambridge Compromise*

The *No-Loss Cambridge Compromise* is a variant of the Cambridge Compromise that incorporates the 2014DP allocation in the form of minimum requirements. For an EP with a total of 705 seats it proceeds as follows:

> Every Member State is allocated five base seats, plus one seat per 870 000 citizens or part thereof, except when the 2014DP restriction warrants more seats or the 96 capping imposes fewer seats.

The resulting allocation is displayed in column "NLCC" of Table 2. The minimum requirement is active from Poland through to Lithuania, while the capping is active for Germany only; these cases are highlighted by a bullet (●). It is straightforward to verify that this allocation obeys degressivity (not shown in Table 2).

### 4.5. *Fix-Proportional-Square-Root Method*

The *Fix-Proportional-Square-Root* method is a variant of the 0.5-DPL method. The method starts by calculating certain indices termed *FPS-scores*. The FPS-score of Member State $i$ is given by the formula

$$3 + 374.4\, p_i + 249.6\, q_i$$

The first number "3" signifies three base seats.[1] This leaves 624 seats yet to be allocated, and these are split into two categories with sizes in proportions 60:40. The second term of the display concerns the first such group of size $60\% \times 624 = 374.4$ seats, which are allocated to States in proportion to population figures.[2] The third term concerns the remaining group of 374.4 seats, which are allocated to States in proportion to the square roots of the population figures.[3] The sum of the three ensuing quantities is called the *FPS-score* of State $i$. The number 3 of base seats is contingent on requirement 8 of Sect. 2.2.

Unlike the 0.5-DPL method, which puts equal weight 0.5 on the proportional and square-root contributions, the Fix-Proportional-Square-Root method uses the weights 0.6 and 0.4.

The FPS-scores having been determined, one now allocates the 705 seats on the basis of the Member States' FPS-scores using the divisor method with standard rounding (Webster method), except when the 2014DP restriction warrants more seats or the 96 capping imposes fewer seats.

---

[1] The base seat total is $27 \times 3 = 81$, thus leaving $705 - 81 = 624$ seats remaining for further consideration.
[2] Here $p_i = P_i/P_+$, where $P_i$ is the QMV2017 population figure of Member State $i$, and $P_+ = P_1 + \ldots + P_{27}$.
[3] Here $q_i = Q_i/Q_+$, where $Q_i$ is the square root of the QMV2017 population figure of Member State $i$, and $Q_+ = Q_1 + \ldots + Q_{27}$.

The resulting seat allocation is exhibited in column "FPS" in Table 2. The minimum requirement is active for Poland, from the Czech Republic through to Bulgaria, and for Lithuania, as indicated by a bullet (●). The 96 capping is dormant. It is straightforward to verify that this allocation obeys degressivity (not shown in Table 2).

### *4.6. The 2019 Composition*

The 2019 composition (column "2019" in Table 2) that emerged finally after lengthy AFCO deliberations appears to be a *deus ex machina*. The allocation satisfies the requirement of degressivity, as verified in the column labelled "RR2019". In other regards it stands isolated. Unlike the two compositions described above, it conforms to neither formula nor methodological procedure. No justification for the composition has been forthcoming from either rapporteurs or speakers as to (i) why the AFCO resolved on a parliament size of 705 seats, and (ii) why the eleven extra seats were allocated to some Member States and denied to others.

It appears to the current authors that the Members of the European Parliament may have overlooked their obligation to justify to the Union citizens the rules which govern their representation. Gray and Stubb (2001) reported from the negotiations in Council during the 2000 Intergovernmental Conference in Nice that, in the final hours, "*the Presidency handed out seats like loose change*". We are left in the dark whether such standards were in operation during the recent AFCO deliberations.

## 5. Conclusion

The AFCO was under an obligation to agree a process which is *objective, fair, durable and transparent* (see Sect. 1). It has not met this obligation, beyond achieving degressivity. There may be some fear in the AFCO that, once a mathematical formula is adopted, it cannot then be varied. This is, of course, false. The United States of America experimented with many different apportionment methods before they decided in 1941, following a century and a half of experience, to codify such a system into law. The AFCO and the EP are overdue on this obligation to democracy.

In its plenary session of 7 February 2018, the EP agreed to the 2019 composition proposed by the AFCO, and forwarded it to the European Council (European Parliament 2018). According to Art. 14(2) TEU, the European Council is asked to adopt the proposition by unanimity. Finally, the EP is expected to give its consent and to close the dossier; see Duff (2018) for more details.

**Table 1: Augmentation of the 2014 composition to achieve degressivity**

|                 | QMV2017     | 2014 | RR2014    | Augm. | 2014DP | RR2014DP |
|-----------------|-------------|------|-----------|-------|--------|----------|
| Germany         | 82 064 489  | 96   | 854 838   | 0     | 96     | 854 838  |
| France          | 66 661 621  | 74   | *900 833  | 4     | 78     | 854 636  |
| Italy           | 61 302 519  | 73   | 839 761   | 0     | 73     | 839 761  |
| Spain           | 46 438 422  | 54   | *859 971  | 2     | 56     | 829 258  |
| Poland          | 37 967 209  | 51   | 744 455   | 0     | 51     | 744 455  |
| Romania         | 19 759 968  | 32   | 617 499   | 0     | 32     | 617 499  |
| The Netherlands | 17 235 349  | 26   | *662 898  | 2     | 28     | 615 548  |
| Belgium         | 11 289 853  | 21   | 537 612   | 0     | 21     | 537 612  |
| Greece          | 10 793 526  | 21   | 513 977   | 0     | 21     | 513 977  |
| Czech Republic  | 10 445 783  | 21   | 497 418   | 0     | 21     | 497 418  |
| Portugal        | 10 341 330  | 21   | 492 444   | 0     | 21     | 492 444  |
| Sweden          | 9 998 000   | 20   | *499 900  | 1     | 21     | 476 095  |
| Hungary         | 9 830 485   | 21   | 468 118   | 0     | 21     | 468 118  |
| Austria         | 8 711 500   | 18   | *483 972  | 1     | 19     | 458 500  |
| Bulgaria        | 7 153 784   | 17   | 420 811   | 0     | 17     | 420 811  |
| Denmark         | 5 700 917   | 13   | *438 532  | 1     | 14     | 407 208  |
| Finland         | 5 465 408   | 13   | †420 416  | 1     | 14     | 390 386  |
| Slovakia        | 5 407 910   | 13   | †415 993  | 1     | 14     | 386 279  |
| Ireland         | 4 664 156   | 11   | *424 014  | 2     | 13     | 358 781  |
| Croatia         | 4 190 669   | 11   | †380 970  | 1     | 12     | 349 222  |
| Lithuania       | 2 888 558   | 11   | 262 596   | 0     | 11     | 262 596  |
| Slovenia        | 2 064 188   | 8    | 258 024   | 0     | 8      | 258 024  |
| Latvia          | 1 968 957   | 8    | 246 120   | 0     | 8      | 246 120  |
| Estonia         | 1 315 944   | 6    | 219 324   | 0     | 6      | 219 324  |
| Cyprus          | 848 319     | 6    | 141 387   | 0     | 6      | 141 387  |
| Luxembourg      | 576 249     | 6    | 96 042    | 0     | 6      | 96 042   |
| Malta           | 434 403     | 6    | 72 401    | 0     | 6      | 72 401   |
| **Sum**         | **445 519 516** | **678** | **—**     | **16** | **694** | **—**    |

**Notes:**

Population figures "QMV2017":
　　Column "QMV2017" contains the population figures that are used for Council's qualified majority voting system during the calendar year 2017, as decreed in European Council (2016).

Columns "2014" and "RR2014":
　　The representation ratio "RR2014" is the quotient of a Member State's "QMV2017" population figure and its status quo "2014" composition, rounded to the nearest whole number. A representation ratio that is larger than its predecessor constitutes a breach of degressivity; it is marked by an asterisk (*) or, when implied by preceding corrections, by a dagger (†).

Columns "2014DP" and "RR2014DP":
　　The "Augm." seats are augmenting the "2014" seats so that the resulting "2014DP" seats achieve degressivity. That is, the representation ratios "RR2014DP" are decreasing when passing from a more populous Member State to a less populous Member State.

**Table 2: NLCC, FPS method, and 2019 composition**

|                 | 2014DP | NLCC | FPS  | 2019 | RR2019  | Increm. |
|-----------------|--------|------|------|------|---------|---------|
| Germany         | 96     | ●96  | 96   | 96   | 854 838 | 0       |
| France          | 78     | 82   | 80   | 79   | 843 818 | 1       |
| Italy           | 73     | 76   | 75   | 76   | 806 612 | 3       |
| Spain           | 56     | 59   | 60   | 59   | 787 092 | 3       |
| Poland          | 51     | ●51  | 51   | 52   | 730 139 | 1       |
| Romania         | 32     | ●32  | ●32  | 33   | 598 787 | 1       |
| The Netherlands | 28     | ●28  | 29   | 29   | 594 322 | 1       |
| Belgium         | 21     | ●21  | 21   | 21   | 537 612 | 0       |
| Greece          | 21     | ●21  | 21   | 21   | 513 977 | 0       |
| Czech Republic  | 21     | ●21  | ●21  | 21   | 497 418 | 0       |
| Portugal        | 21     | ●21  | ●21  | 21   | 492 444 | 0       |
| Sweden          | 21     | ●21  | ●21  | 21   | 476 095 | 0       |
| Hungary         | 21     | ●21  | ●21  | 21   | 468 118 | 0       |
| Austria         | 19     | ●19  | ●19  | 19   | 458 500 | 0       |
| Bulgaria        | 17     | ●17  | ●17  | 17   | 420 811 | 0       |
| Denmark         | 14     | ●14  | 14   | 14   | 407 208 | 0       |
| Finland         | 14     | ●14  | 14   | 14   | 390 386 | 0       |
| Slovakia        | 14     | ●14  | 14   | 14   | 386 279 | 0       |
| Ireland         | 13     | ●13  | 13   | 13   | 358 781 | 0       |
| Croatia         | 12     | ●12  | 12   | 12   | 349 222 | 0       |
| Lithuania       | 11     | ●11  | ●11  | 11   | 262 596 | 0       |
| Slovenia        | 8      | 8    | 9    | 8    | 258 024 | 0       |
| Latvia          | 8      | 8    | 8    | 8    | 246 120 | 0       |
| Estonia         | 6      | 7    | 7    | 7    | 187 992 | 1       |
| Cyprus          | 6      | 6    | 6    | 6    | 141 387 | 0       |
| Luxembourg      | 6      | 6    | 6    | 6    | 96 042  | 0       |
| Malta           | 6      | 6    | ●6   | 6    | 72 401  | 0       |
| **Sum**         | **694**| **705** | **705** | **705** | —   | **11**  |

**Notes:**

No-Loss Cambridge Compromise "NLCC":
   Every Member State is allocated 5 base seats, plus one seat per 870 000 citizens or part thereof, except when the 2014DP-restriction warrants more seats or the 96 capping imposes fewer seats; exceptions are marked by a bullet (●).

Fix-Proportional-Square-Root method "FPS":
   Seats are apportioned proportionately to FPS-scores using the divisor method with standard rounding (Webster method), except when the 2014DP-restriction warrants more seats; exceptions are marked by a bullet (●). Member State $i$ has FPS-score $3 + 374.4 p_i + 249.6 q_i$, with $p_i = P_i/P_+$ and $q_i = Q_i/Q_+$, where $P_i$ is the QMV-2017 population figure of state $i$ and $Q_i$ is its square root, while $P_+ = P_1 + \ldots + P_{27}$ and $Q_+ = Q_1 + \ldots + Q_{27}$.

Columns "2019" and "RR2019"
   Column "2019" shows the 2019–2024 EP composition adopted by AFCO and the EP. Column "RR2019" verifies that representation ratios are decreasing when passing from a more populous Member State to a less populous Member State.

Column "Increm." exhibits the increments from "2014DP" seats to "2019" seats.

# References

DIRECTORATE-GENERAL FOR INTERNAL POLICIES, POLICY DEPARTMENT C: CITIZENS' RIGHTS AND CONSTITUTIONAL AFFAIRS, AND COMMITTEE ON CONSTITUTIONAL AFFAIRS OF THE EUROPEAN PARLIAMENT. 2017. *The Composition of the European Parliament. Workshop 30 January 2017.* PE 583.117, February 2017 (www.uni-augsburg.de/pukelsheim/2017Brussels/).

DUFF, ANDREW. 2018. How to Govern Europe better: Reflections on Reform of the European Parliament, Commission and Council. European Policy Centre, Working Paper 13 February 2018 (www.epc.eu/pub_details.php?cat_id=17&pub_id=8271).

EUROPEAN COUNCIL. 2013. Decision of 28 June 2013 establishing the composition of the European Parliament (2013/312/EU). *Official Journal of the European Union* L 181, 29.6.2013: 57–8 (www.uni-augsburg.de/bazi/OJ/2013L181p57.pdf).

EUROPEAN COUNCIL. 2016. Decision (EU, Euratom) 2016/2353 of 8 December 2016 amending the Council's Rules of Procedure. *Official Journal of the European Union* L 348, 12.12.2016: 27–9 (www.uni-augsburg.de/bazi/OJ/2016L348p27.pdf).

EUROPEAN PARLIAMENT. 2008. Composition of the European Parliament – Resolution of 11 October 2007 on the composition of the European Parliament (2007/2169(INI)). *Official Journal of the European Union* C 227 E, 4.9.2008: 132–8 (www.uni-augsburg.de/bazi/OJ/2008C227Ep132.pdf).

EUROPEAN PARLIAMENT. 2018. Composition of the European Parliament – Resolution of 7 February 2018 on the composition of the European Parliament (2017/2054(INL)-2017/0900(NLE)). (www.uni-augsburg.de/pukelsheim/2017Brussels/).

EUROPEAN UNION. 2012. Consolidated version of the Treaty on European Union. *Official Journal of the European Union* C 326, 26.10.2012: 13–45 (www.uni-augsburg.de/bazi/OJ/2012C326p13.pdf).

GRAY, MARK and ALEXANDER STUBB. 2001. Keynote article: The Treaty of Nice – Negotiating a Poisoned Chalice? *Journal of Common Market Studies* 39: 12–23.

GRIMMETT, GEOFFREY, JEAN-FRANÇOIS LASLIER, FRIEDRICH PUKELSHEIM, VICTORIANO RAMÍREZ GONZÁLEZ, RICHARD ROSE, WOJCIECH SŁOMCZYŃSKI, MARTIN ZACHARIASEN and KAROL ŻYCZKOWSKI. 2011. *The Allocation Between the EU Member States of the Seats in the European Parliament – Cambridge Compromise*. Note. European Parliament, Directorate-General for Internal Policies, Policy Department C: Citizen's Rights and Constitutional Affairs, PE 432.760, March 2011 (www.uni-augsburg.de/pukelsheim/2011f.pdf).

GRIMMETT, GEOFFREY, KAI-FRIEDERIKE OELBERMANN and FRIEDRICH PUKELSHEIM. 2012. A power-weighted variant of the EU27 Cambridge Compromise. *Mathematical Social Sciences* 63: 136–40 (www.uni-augsburg.de/pukelsheim/2012a.pdf).

GRIMMETT, GEOFFREY, FRIEDRICH PUKELSHEIM, VICTORIANO RAMÍREZ GONZÁLEZ, WOJCIECH SŁOMCZYŃSKI, and KAROL ŻYCZKOWSKI. 2017. *A 700-seat no-loss composition for the 2019 European Parliament.* 22 August 2017 (www.uni-augsburg.de/pukelsheim/2017Brussels/).

HÜBNER, DANUTA MARIA and PEDRO SILVA PEREIRA. 2018. *Report on the Composition of the European Parliament (2017/2054(INL) - 2017/0900(NLE)).* Committee on Constitutional Affairs of the European Parliament. PE 608.038v02-00, 26 January 2018 (www.uni-augsburg.de/pukelsheim/2017Brussels/).

PUKELSHEIM, FRIEDRICH. 2017. *Proportional Representation. Apportionment Methods and Their Applications. With a Foreword by Andrew Duff MEP. Second Edition.* Cham: Springer International Publishing, 2017 (www.uni-augsburg.de/pukelsheim/2017a-FrontMatter.pdf).

RAMÍREZ GONZÁLEZ, VICTORIANO. 2017. *Composition of the European Parliament – The FPS-Method.* Typescript, 20 August 2017 (www.uni-augsburg.de/pukelsheim/2017Brussels/).

RAMÍREZ, VICTORIANO, ANTONIO PALOMARES and MARIA LUISA MÁRQUEZ. 2006. Degressively proportional methods for the allotment of the European Parliament Seats amongst the EU Member States. In: SIMEONE, BRUNO and FRIEDRICH PUKELSHEIM (eds.): *Mathematics and Democracy – Recent Advances in Voting Systems and Collective Choice*, Berlin: Springer, pp. 205–20.

SŁOMCZYŃSKI, WOJCIECH and KAROL ŻYCZKOWSKI. 2012. Mathematical aspects of degressive proportionality. *Mathematical Social Sciences* 63: 94–101.